# Hadronic Cross Sections, Elastic Slope and Physical Bounds

D.A. Fagundes and M.J. Menon

*Instituto de Física Gleb Wataghin, Universidade Estadual de Campinas, UNICAMP, 13083-859 Campinas SP, Brazil*

**Abstract.** An almost model-independent parametrization for the ratio of the total hadronic cross section to elastic slope is discussed. Its applicability in studies of asymptotia and analyses of extensive air shower in cosmic-ray physics is also outlined.

**Keywords:** Hadron-induced high- and super-high-energy interactions, Cosmic-ray interactions
**PACS:** 13.85.-t, 13.85.Tp

In cosmic-ray physics, extensive air showers (EAS) constitute an important tool for analyzing particle and nuclear physics at energies far beyond those obtained in accelerator machines. However, the physical interpretation of the EAS relies on extrapolations from phenomenological models solely tested in the accelerator energy region, hence giving rise to systematic theoretical uncertainties. One model-dependent quantity demanding extrapolation involves the total cross section, $\sigma_{tot}$, and the forward elastic slope, $B_{el}$. Here we introduce a novel analytical parameterization for the ratio $\sigma_{tot}/B_{el}(s)$, considering its connection with the ratio of the elastic to total cross section, $R(s)$, through which experimental data is fitted.

On deriving model-independent results, general principles and high-energy theorems constitute essential tools for providing constraints to the asymptotic energy regime. In this context, we exploit here the formal bound by MacDowell and Martin [1],

$$\frac{\sigma_{tot}(s)}{B_{el}(s)} \leq 18\pi \frac{\sigma_{el}(s)}{\sigma_{tot}(s)}, \qquad (1)$$

as discussed in [2].
In addition, experimental data provide valuable trends on asymptotia. Specially, we shall focus on the experimental data on elastic, $\sigma_{el}(s)$, and total cross sections, $\sigma_{tot}(s)$, but also on the forward elastic slope, $B_{el}$, which is defined by the parameterization of the differential elastic cross section at the diffraction cone:

$$\frac{d\sigma}{dt} = \left.\frac{d\sigma}{dt}\right|_{t=0} e^{B_{el}t}, \qquad (2)$$

Integrating the last equation over the $t$ range $[-\infty,0]$, we get [2]:

$$\frac{\sigma_{tot}(s)}{B_{el}(s)} = 16\pi\frac{\sigma_{el}(s)}{\sigma_{tot}(s)} \equiv 16\pi R(s). \qquad (3)$$

The above result allows one to investigate the energy dependence of $\sigma_{tot}/B_{el}(s)$ from fits to the experimental data on the ratio $R(s)$. Although the experimental data on the ratio $R$ do not allow us to antecipate its asymptotic behaviour, two possibilities for asymptotia are predicted:

(i) the black disk limit is attained, i.e. $R(s) \to A = 1/2$ [3];
(ii) the obvious unitarity bound is saturated, i.e $R(s) \to A = 1$ [4].

Regarding both possible scenarios, we have performed tests on goodness of fit to the $R$ data. The best statistical result was obtained with the model-independent parametrization:

$$R(s) \equiv \frac{\sigma_{el}}{\sigma_{tot}}(s) = A\tanh(\gamma_1 + \gamma_2 \ln s + \gamma_3 \ln^2 s), \qquad (4)$$

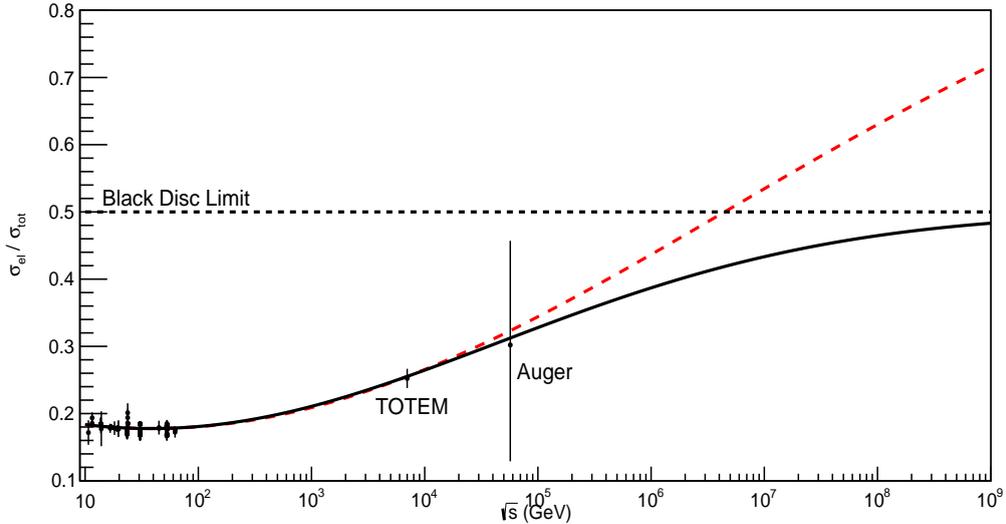

**FIGURE 1.** Fit results for $R(s)$ in Eq. (3) with $A = 1$ (dashed line) and $1/2$ (solid line). In both fits $DOF = 87$ and $\chi^2/DOF = 1.2$.

From eq. (3), the energy dependence of $\sigma_{tot}/B_{el}$ follows straightforwardly. In Fig. 2 we display $\sigma_{tot}/B_{el}(s)$ predicted from (3) and in Table 1 we present our numerical predictions for the same ratio at the LHC energies.

From these results, we drive the following main conclusions:

1. the black disk behaviour is only achieved at energies $\sqrt{s} \gtrsim 10^9$ GeV;

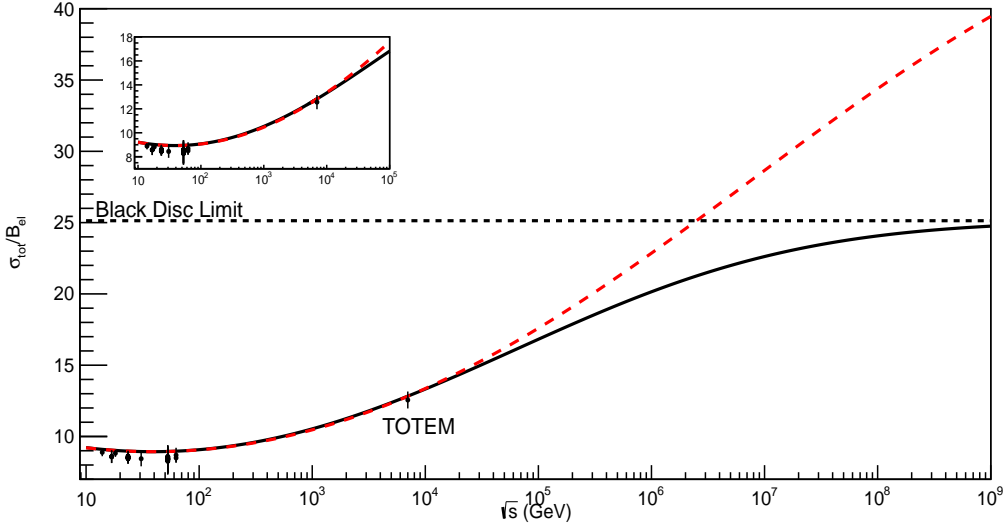

**FIGURE 2.** Experimental data on the ratio $\sigma_{tot}/B_{el}$ [5] and predictions from eq. (3) for $A = 1$ (dashed line) and $1/2$ (solid line). The inset displays a zoom of the figure up to the Auger energy region.

**TABLE 1.** Predictions from eq. (3) for the ratio $\sigma_{tot}/B_{el}$ at the LHC energy region and the TOTEM result at 7 TeV [5].

| $\sqrt{s}$ | A = 1/2 | A = 1 | TOTEM |
|---|---|---|---|
| 7.0 TeV | 12.827±0.047 | 12.821±0.024 | 12.56±0.59 |
| 14 TeV | 13.811±0.068 | 13.903±0.033 | – |

2. present experimental information may indicate an antishadow contribution at $\sqrt{s} \gtrsim 2 \times 10^6$ GeV;
3. regarding $\sigma_{tot}(s) \sim \ln^2(s)$, $B_{el}(s)$ acquires a $\ln^2(s)$ component, in accordance with the effective rise with energy of $\alpha_{\mathbb{P}}^{'eff}$ [6];
4. even in the extrema cases, $A = 1$ or $A = 1/2$, the ratio $R(s = (57\,\text{TeV})^2) = 0.30^{+0.17}_{-0.16}$, following from the recent estimations of inelastic and total cross section by the Pierre Auger Collaboration [7], is well described.
5. extrapolations of $\sigma_{tot}/B_{el}(s)$ to Auger energy region stand in the range $15.5 - 16.3$.

**Acknowledgments:** Research supported by FAPESP (contracts 11/00505-0 and 09/50180-0). We are thankful to P.V.R.G. Silva for discussions.